\documentclass[slac_one]{revtex4}
\usepackage{graphicx,fancyhdr,amsmath,amssymb,colordvi,pifont}
\allowdisplaybreaks

\newcommand\eg{e.g.\ }
\newcommand\ie{i.e.\ }

\newcommand\ord{\mathcal{O}}
\newcommand\ri{\mathrm{i}}
\newcommand\re{\mathrm{e}}

\newcommand\nbsp{\hspace*{2em}}
\newcommand\twoloop{\ding{192}}
\newcommand\oneloop{\ding{193}}
\newcommand\qdep{\ding{194}}

\begin{document}

\title{{\small{2005 International Linear Collider Workshop - Stanford,
U.S.A.}}\\ 
\vspace{12pt}
Precision Higgs Masses with FeynHiggs 2.2} 

\author{Thomas Hahn, Wolfgang Hollik}
\affiliation{Max-Planck-Institut f\"ur Physik,
        F\"ohringer Ring 6,
        80805 Munich, Germany}

\author{Sven Heinemeyer}
\affiliation{CERN TH division,
        Dept.\ of Physics,
        1211 Geneva 23, Switzerland}

\author{Georg Weiglein}
\affiliation{Institute for Particle Physics Phenomenology,
        University of Durham,
        Durham DH1~3LE, UK}

\begin{abstract}
FeynHiggs is a program for computing MSSM Higgs-boson masses and related
observables, such as mixing angles, branching ratios, and couplings, 
including state-of-the-art higher-order contributions.  The centerpiece 
is a Fortran library for use with Fortran and C/C++.  Alternatively, 
FeynHiggs has a command-line, Mathematica, and Web interface.  The 
command-line interface can process, besides its native format, files in 
SUSY Les Houches Accord format.  FeynHiggs is an open-source program and 
easy to install.
\end{abstract}

\maketitle

\thispagestyle{fancy}


\section{Introduction}

One of main goals of future colliders is to find a Higgs boson. In order
to establish the mechanism of electroweak symmetry breaking it will in
addition be necessary to measure the properties of the Higgs boson,
hopefully allowing to distinguish between different models.  While the
LHC will almost certainly take the prize of finding a Higgs
\cite{LHCHiggs}, if it exists, it will take the International
Linear Collider (ILC) to nail down many of the properties
\cite{tesla,orangebook,acfarep} to the desired level of accuracy.

Unlike in the Standard Model (SM), where the Higgs mass is only rather
loosely constrained by higher-order effects, the Higgs couplings in the
Minimal Supersymmetric Standard Model (MSSM) \cite{susy} are directly
related, through supersymmetry, to the gauge couplings.  This implies
that the lightest Higgs-boson mass $M_h$ can be predicted in terms of
the other model parameters.  The mass measurement at the ILC is
estimated to $\delta M_h^{\text{exp}} \approx 0.05$ GeV
\cite{tesla,orangebook,acfarep}, thus $M_h$ will become a precision
observable.

Together, these two issues mandate precise calculations of observables
on the theory side in a variety of models, but in particular in
supersymmetric ones, where $M_h$ is a prediction.  The FeynHiggs
\cite{FeynHiggs,FeynHiggs2.2} package provides masses, couplings,
branching ratios, etc.\ in the real, complex, and non-minimal
flavour-violating MSSM including state-of-the-art radiative corrections.


\section{The MSSM Higgs sector}

The MSSM contains two Higgs doublets,
\begin{equation}
H_1 = \begin{pmatrix}
      v_1 + \frac 1{\sqrt 2}(\phi_1 + \ri\chi_1) \\
      \phi_1^-
      \end{pmatrix},
\qquad
H_2 = \Magenta{\re^{\ri\xi}}
      \begin{pmatrix}
      \phi_2^+ \\
      v_2 + \frac 1{\sqrt 2}(\phi_2 + \ri\chi_2)
      \end{pmatrix},
\end{equation}
where a possible CP-violating phase $\xi$ has been indicated.  The Higgs 
potential is given by
\begin{equation}
V = m_1^2 H_1\bar H_1 + m_2^2 H_2\bar H_2 -
    \Magenta{m_{12}^2} (\varepsilon_{\alpha\beta}
      H_1^\alpha H_2^\beta + \mathrm{h.c.}) +
    \frac{g_1^2 + g_2^2}{8}\,(H_1\bar H_1 - H_2\bar H_2)^2 +
    \frac{g_2^2}{2}\,|H_1\bar H_2|^2.
\end{equation}
The only non-trivial CP-violating phase (besides $\xi$) is contained
here in $m_{12}$.  At tree level all CP phases can be rotated away,
giving five physical states of distinct CP parity: $h^0$, $H^0$
(CP-even), $A^0$ (CP-odd), and $H^\pm$.

The quartic Higgs couplings are completely determined by the gauge
couplings $g_1$ and $g_2$ and this leads to the well-known tree-level
prediction $M_h < M_Z$, which stands in conflict with measurements since
LEP.  Fortunately (for the MSSM), significant quantum loop contributions
push the upper bound on $M_h$ up to about 140 GeV
\cite{mhiggslong,mhiggsAEC} for a top mass of 178 GeV.  But the quantum
effects lead also to qualitative changes. In the presence of
CP-violating phases, all three neutral Higgs bosons mix and CP is no
longer conserved \cite{mhiggsCPV},
\begin{equation}
\begin{pmatrix}
h_1 \\
h_2 \\
h_3
\end{pmatrix} = \begin{pmatrix}
U_{11} & U_{12} & \Magenta{U_{13}} \\
U_{21} & U_{22} & \Magenta{U_{23}} \\
\Magenta{U_{31}} & \Magenta{U_{32}} & U_{33}
\end{pmatrix} \begin{pmatrix}
h^0 \\
H^0 \\
A^0
\end{pmatrix}.
\end{equation}
The three mass eigenstates are denoted as $h_i$, $i = 1,2,3$, and
ordered as $m_{h_1} \leqslant m_{h_2} \leqslant m_{h_3}$.


\section{FeynHiggs}

\subsection{Download and Installation}

Installing FeynHiggs is simple and fast.  Version 2.2 requires no 
prerequisites (\eg LoopTools) as before.
\begin{itemize}
\itemsep=0pt
\item
Get the latest FeynHiggs tar file from \texttt{http://www.feynhiggs.de}.

\item
Unpack, configure, and build: \\
\nbsp\texttt{tar xfz FeynHiggs-2.2.$N$.tar.gz} \\
\nbsp\texttt{cd FeynHiggs-2.2.$N$} \\
\nbsp\texttt{./configure} \\
\nbsp\texttt{make} \\
To build also the Mathematica part, replace ``\texttt{make}'' by
``\texttt{make all}''.

\item
(Optional:) Type ``\texttt{make install}'' to install the files and
``\texttt{make clean}'' to remove intermediate files.
\end{itemize}


\subsection{Modes of Operation}

FeynHiggs operates in one of four basic modes:
\begin{itemize}
\itemsep=0pt
\item
Library Mode: The FeynHiggs routines are invoked from a Fortran or C/C++ 
program linked against the FeynHiggs library.

\item
Command-line Mode: Parameter files in FeynHiggs' native format or in
SUSY Les Houches Accord (SLHA) \cite{SLHA} format are processed at the
command line 
with the standalone executable \texttt{FeynHiggs}.

\item
WWW Mode: The user interactively chooses parameters at the FeynHiggs 
User Control Center (FHUCC) and obtains the results on-line
at \texttt{http://www.feynhiggs.de/fhucc} .

\item
Mathematica Mode: The FeynHiggs routines can be used in Mathematica via
the MathLink executable \texttt{MFeynHiggs}.

\end{itemize}


\subsection{Application Programming Interface}
\label{sect:api}

The FeynHiggs library \texttt{libFH.a} is a static Fortran 77 library. 
Its global symbols are prefixed with a unique identifier to minimize
symbol collisions.  The library contains only subroutines (no
functions), so that no include files are needed (except for the
couplings) and the invocation from C/C++ is hassle-free.  Detailed
debugging output can be turned on at run time.  All routines are 
described in detail in the API guide and on man-pages, so only a brief 
overview is needed here:
\begin{itemize}
\itemsep=0pt
\item
\texttt{FHSetFlags} sets the flags of the calculation.

\item
\texttt{FHSetPara} sets the MSSM input parameters directly.

\item
\texttt{FHSetSLHA} extracts the input parameters from an SLHA data
structure.

\item
\texttt{FHSetDebug} sets the debugging level.

\item
\texttt{FHGetPara} retrieves (some of) the derived parameters.

\item
\texttt{FHHiggsCorr} evaluates the Higgs masses and mixings, 
$M_{h_1,h_2,h_3,H^\pm}$, $\alpha_{\mathrm{eff}}$ (the effective mixing 
angle in the CP-conserving case), $U_{ij}$, featuring:
\begin{itemize}
\item
In the neutral Higgs sector, the following propagator matrix is 
diagonalized,
\begin{equation}
\begin{pmatrix}
q^2 - M_h^2 + \hat\Sigma_{hh}^{\text{\twoloop\oneloop\qdep}} &
        \hat\Sigma_{hH}^{\text{\twoloop\oneloop\qdep}} &
                \Magenta{\hat\Sigma_{hA}^{\text{\oneloop\qdep}}}
\\[.4ex]
\hat\Sigma_{Hh}^{\text{\twoloop\oneloop\qdep}} &
        q^2 - M_H^2 + \hat\Sigma_{HH}^{\text{\twoloop\oneloop\qdep}} &
                \Magenta{\hat\Sigma_{HA}^{\text{\oneloop\qdep}}}
\\[.4ex]
\Magenta{\hat\Sigma_{Ah}^{\text{\oneloop\qdep}}} &
        \Magenta{\hat\Sigma_{AH}^{\text{\oneloop\qdep}}} &
                q^2 - M_A^2 + \hat\Sigma_{AA}^{\text{\oneloop\qdep}}
\end{pmatrix},
\end{equation}
where the self-energies include the following terms as indicated,
\begin{itemize}
\item[\twoloop]
the most up-to-date leading $\ord(\alpha_s\alpha_t, \alpha_t^2)$
\cite{mhiggslong,asat,atat} and subleading $\ord(\alpha_s\alpha_b, 
\alpha_t\alpha_b, \alpha_b^2)$ \cite{asab,atab} two-loop corrections in 
the rMSSM (complex effects are taken into account only partially in the
two-loop part at present),

\item[\oneloop]
full one-loop evaluation (all phases included),

\item[\qdep]
complete $q^2$ dependence.
\end{itemize}

\item
Full one-loop corrections for the charged Higgs sector \cite{MH1loop}.

\item
Mixed $\overline{\text{DR}}$/OS renormalization for the one-loop result 
\cite{MSbarOS}.

\item
``$\Delta m_b$'' corrections = leading $\ord(\alpha_s\alpha_b)$ and 
$\ord(\alpha_t\alpha_b)$ terms for Higgs masses, couplings, etc.\ 
\cite{DeltaMB}.

\item
Non-minimal flavour-violating effects
(\eg $\tilde c$--$\tilde t$ mixing) \cite{NMFV}.
\end{itemize}

\item
\texttt{FHUncertainties} estimates the uncertainties of the Higgs 
masses and mixings.  The total uncertainty is the sum of
deviations from the central value, $\Delta X = \sum_{i = 1}^3
|X_i - X|$ with $X = \{M_{h_1,h_2,h_3,H^\pm}, \alpha_{\mathrm{eff}}, 
U_{ij}\}$, where
\begin{itemize}
\item
$X_1$ is obtained by varying the renormalization scale (entering via
the $\overline{\text{DR}}$ renormalization) within
$\frac 12 m_t \leqslant \mu \leqslant 2 m_t$,

\item
$X_2$ is obtained by using $m_t^{\text{pole}}$ instead of the running 
$m_t$ in the two-loop corrections,

\item
$X_3$ is obtained by using an unresummed bottom Yukawa coupling,
$y_b$, \ie an $y_b$ including  
the leading $\ord(\alpha_s\alpha_b)$ corrections, but not resummed to 
all orders.
\end{itemize}

\item
\texttt{FHCouplings} computes the Higgs couplings, decay widths, and BRs 
in the MSSM and also for an SM Higgs boson with mass $M_{h_i}$
(denoted as $h_{1,2,3}^{\text{SM}}$) for
comparison:
\begin{equation}
\begin{aligned}
h_{1,2,3} \to {}
& f\bar f, \gamma\gamma, ZZ^*, WW^*, gg, \qquad\qquad &
        H^\pm \to {}
        & f\bar f', &
                h_{1,2,3}^{\text{SM}} \to {}
                & f\bar f, \gamma\gamma, ZZ^*, WW^*, gg. \\
& h_i Z^*, h_i h_j, H^+ H^-, &
        & h_i W^{\pm *}, \qquad\qquad \\
& \tilde f_i \tilde f_j, &
        & \tilde f_i \tilde f'_j, \\
& \tilde\chi_i^\pm \tilde\chi_j^\pm, \tilde\chi_i^0 \tilde\chi_j^0, &
        & \tilde\chi_i^0 \tilde\chi_j^\pm,
\end{aligned}
\end{equation}

\item
\texttt{FHHiggsProd} calculates approximately (by means of effective 
couplings) the following inclusive Higgs production cross-sections:
$bb \to h + X$, $gg \to h + X$, $qq \to qqh + X$, $qq,gg \to tth + X$,
$qq \to Wh + X$, $qq \to Zh + X$ \cite{HiggsXS}.

\item
\texttt{FHConstraints} evaluates several electroweak precision
observables, to be used as additional constraints:
\begin{itemize}
\item
$\Delta\rho$ at $\ord(\alpha, \alpha\alpha_s)$
\cite{delrhosusy2loop,PomssmRep}.  Too large values of $\Delta\rho$
indicate experimentally disfavoured $\tilde t$/$\tilde b$ masses.

\item
$(g_\mu - 2)_{\mathrm{SUSY}}$ including full one-loop and
leading/subleading two-loop SUSY corrections \cite{g-2,g-2CNH}.

\item
(Preliminary:) The electric dipole moments of Th, N, and Hg.  This 
part is not yet fully tested.
\end{itemize}
\end{itemize}


\section{Command-line Modes}

The FeynHiggs command-line frontend, \texttt{FeynHiggs}, reads input 
files both in its own and in SLHA format.  The FeynHiggs format simply 
lists the parameters and their values, for example:
\begin{verbatim}
  MT         178
  MB         4.7
  MW         80.450
  MZ         91.1875
  TB         50
  MA0        200
  MSusy      975
  ...
\end{verbatim}
More sophisticated variants are possible, \eg ``\texttt{TB 5 50 2.5}''
declares a loop over \texttt{TB} = $\tan\beta$ from 5 to 50 in steps of 
2.5.  This input file, e.g.\ \texttt{fh.in}, is run through FeynHiggs by
\begin{verbatim}
  FeynHiggs fh.in
\end{verbatim}
Optionally, the flags can be given behind the filename as a string of 
digits, as in
\begin{verbatim}
  FeynHiggs fh.in 40020211
\end{verbatim}
The output is listed on stdout in a human-readable form, for example
\begin{verbatim}
--------------------- HIGGS MASSES ---------------------
| Mh0         =     117.186672
| MHH         =     194.268239
| MA0         =     200.000000
| MHp         =     212.662071
| SAeff       =    -0.36496659
| UHiggs      =     0.99589960    0.09046538    0.00000000 \
|                  -0.09046538    0.99589960    0.00000000 \
|                   0.00000000    0.00000000    1.00000000

---------------- ESTIMATED UNCERTAINTIES ---------------
| DeltaMh0    =       0.919435
| DeltaMHH    =       0.728304
| DeltaMA0    =       0.000000
| DeltaMHp    =       1.929728
  ...
\end{verbatim}
The listing can become quite lengthy, and although FeynHiggs 
automatically spawns a pager for easier viewing, one would sometimes 
like to mask off the details.  Such lines contain a \% character, thus
\begin{verbatim}
   FeynHiggs fh.in | grep -v %
\end{verbatim}
turns off the details.

To convert the human-readable into a machine-readable form, the
\texttt{table} utility is used.  For example, the following line
produces a file \texttt{fh.out} with two columns, \texttt{TB} and
\texttt{Mh0},
\begin{verbatim}
   FeynHiggs fh.in | table TB Mh0 > fh.out
\end{verbatim}

The SLHA mode works similarly, only that the output is not listed on 
screen, but saved in a file (input filename plus ``\texttt{.fh}''), 
again in SLHA format.  This way, FeynHiggs can act as a filter in a 
chain of commands operating on an SLHA file.  FeynHiggs tries to read 
each input file in SLHA format first and if that fails, falls back into 
its native format.  FeynHiggs' SLHA interface uses the SLHA Library 
\cite{SLHALib}.


\section{Interactive Modes}

\begin{figure}
\includegraphics{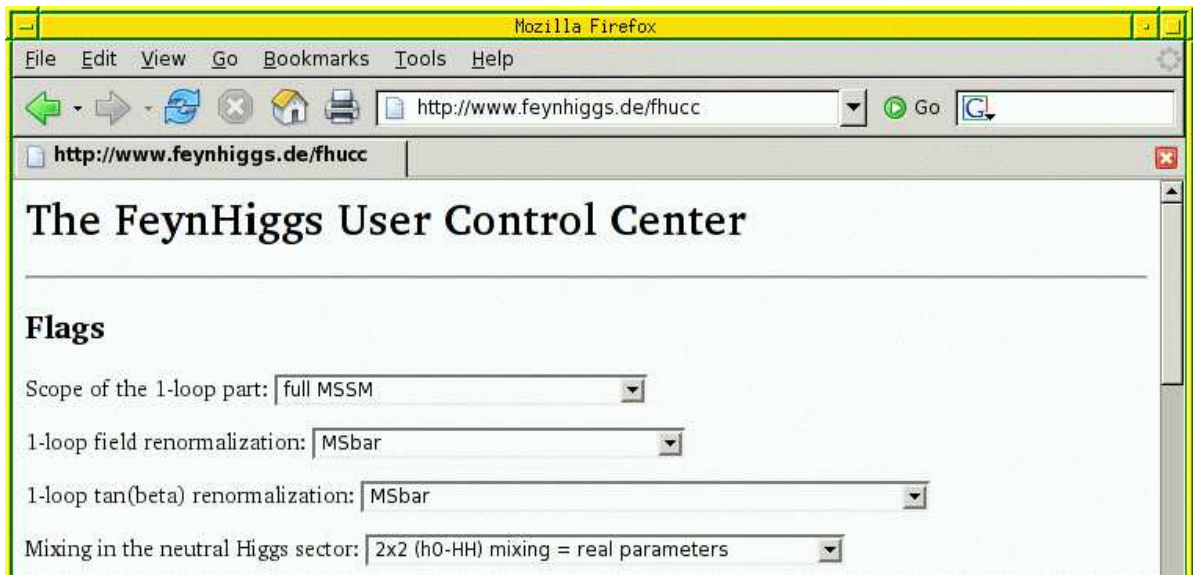}
\caption{\label{fig:fhucc}The FeynHiggs User Control Center.}
\end{figure}

FeynHiggs can be used interactively in WWW Mode or in Mathematica Mode.
In WWW Mode, point your browser to the FeynHiggs User Control Center at
\texttt{http://www.feynhiggs.de/fhucc}.  The Web interface allows to 
select one of the Les Houches benchmark scenarios, or choose each 
parameter directly.  Fig.\ \ref{fig:fhucc} shows a screen shot.

A much more powerful interactive environment is provided by the 
Mathematica interface of FeynHiggs.  The MathLink executable 
\texttt{MFeynHiggs} must first be loaded with
\begin{verbatim}
   Install["MFeynHiggs"]
\end{verbatim}
and makes all FeynHiggs routines (see Sect.\ \ref{sect:api}) available 
as Mathematica functions.  Standard Mathematica functions, such as
\texttt{ContourPlot} and \texttt{FindMinimum}, then make some 
sophisticated analyses possible.


\section{Summary}

The FeynHiggs package computes Higgs masses, mixing angles, branching
ratios, couplings, etc.\ in the MSSM including state-of-the-art
radiative corrections.  The heart of the program is a static Fortran
library which can be accessed either directly (in Fortran or C/C++) or
through various frontends (command-line, Mathematica, WWW).  FeynHiggs
is freely available from \texttt{http://www.feynhiggs.de} and is
straightforward to compile and install.


\begin{flushleft}

\end{flushleft}

\end{document}